# Superconductivity-driven ferromagnetism and spin manipulation using vortices in the magnetic superconductor EuRbFe$_4$As$_4$

Shigeyuki Ishida[1*], Daniel Kagerbauer[2], Sigrid Holleis[2], Kazuki Iida[3], Koji Munakata[3], Akiko Nakao[3], Akira Iyo[1], Hiraku Ogino[1], Kenji Kawashima[4], Michael Eisterer[2] and Hiroshi Eisaki[1].

[1] Research Institute for Advanced Electronics and Photonics, National Institute of Advanced Industrial Science and Technology, Tsukuba, Ibaraki 305-8568, Japan

[2] Atominstitut, TU Wien, Stadionallee 2, 1020 Vienna, Austria

[3] Neutron Science and Technology Center, Comprehensive Research Organization for Science and Society (CROSS), Tokai, Ibaraki 319-1106, Japan

[4] Research & Development Department, IMRA JAPAN CO., LTD., Kariya, Aichi 448-8650, Japan

*Shigeyuki Ishida

Email:  s.ishida@aist.go.jp

**Magnetic superconductors are specific materials exhibiting two antagonistic phenomena, superconductivity and magnetism, whose mutual interaction induces various emergent phenomena, such as the re-entrant superconducting transition associated with the suppression of superconductivity around the magnetic transition temperature ($T_m$), highlighting the impact of magnetism on superconductivity. In this study, we report the experimental observation of the ferromagnetic order induced by superconducting vortices in the high-critical-temperature (high-$T_c$) magnetic superconductor EuRbFe$_4$As$_4$. Although the ground state of the Eu$^{2+}$ moments in EuRbFe$_4$As$_4$ is helimagnetism below $T_m$, neutron diffraction and magnetization experiments show a ferromagnetic hysteresis of the Eu$^{2+}$ spin alignment. We demonstrate that the direction of the Eu$^{2+}$ moments is dominated by the distribution of pinned vortices based on the critical state model. Moreover, we demonstrate the manipulation of spin texture by controlling the direction of superconducting vortices, which can help realize spin manipulation devices using magnetic superconductors.**

## Introduction

The coexistence of superconductivity and magnetism has been a long-standing issue in the field of superconductivity due to the antagonistic nature of these two ordered states. Therefore, magnetic superconductors, exhibiting both behaviours simultaneously, have played an important role in the study of their interaction. Various novel phenomena have been reported in magnetic superconductors containing rare earth elements ($R$) such as Ho and Er, e.g., $R$Rh$_4$B$_4$ ($T_c$ ~ 9 K and $T_m$ ~ 1 K) (1), $R$Mo$_6$S$_8$ ($T_c$ ~ 2 K and $T_m$ ~ 0.2-1 K) (2), and $R$Ni$_2$B$_2$C ($T_c$ ~ 5-10 K and $T_m$ ~ 2-6 K) (3, 4), including re-entrant superconducting transition, anomalous temperature dependence of the upper critical field, and enhanced vortex pinning (or critical current density ($J_c$)) associated with the magnetic transition. All have contributed significantly to the development of superconductivity research.

Recently, iron-based superconductors containing Eu (Eu-IBSs) have attracted attention as a new class of magnetic superconductors (5). Eu-IBSs are characterized by higher $T_c$ (up to 37 K) and $T_m$ (~15-20 K) compared to other magnetic superconductors. The coexistence of superconductivity and magnetism extends over a wider range of temperatures and magnetic fields, allowing us to conduct experiments using various probes. For example, in the case of EuFe$_2$(As,P)$_2$ (6) and Eu(Fe,Rh)$_2$As$_2$ (7), where the Eu$^{2+}$ moments exhibit ferromagnetic ordering, observations of a spontaneous vortex state, the domain Meissner state, and a vortex-antivortex state were reported using magnetic force microscopy (8) and magnetization measurements (9). Furthermore, for EuRbFe$_4$As$_4$ (10, 11) with a Eu$^{2+}$ helical order (12), it was proposed that the Eu-spin subsystem serves as an internal pump for the magnetic flux based on magneto-optical imaging results (13, 14). In addition, optical conductivity measurements (15) and scanning Hall microscopy (16) revealed that the temperature dependence of the superfluid density shows a dip feature around $T_m$, which was attributed to the weakening of superconductivity at the magnetic transition.

Most of these peculiar phenomena in magnetic superconductors are considered to reflect the impact of magnetism on superconductivity. On the contrary, an example of the influence of superconductivity on magnetism is the shrinkage of magnetic domains associated with the superconducting transition (17), which is favorable for superconductivity, as it reduces the internal magnetic field. Meanwhile, in a type-II superconductor, the magnetic flux penetrates the material and the mixed state is generated. Once vortices are pinned, the superconductor can act as a magnet (known as a trapped-field magnet). In this situation, magnetism is likely influenced by the magnetic field generated by superconductivity. For example, the manipulation of local moments via superconducting vortices has been proposed using superconductor-magnet hybrids (18). To further investigate the impact of pinned vortices on the local moments, a superconductor with strong pinning properties is considered as a suitable target.

The IBSs, especially the 122, 1111 and 1144 types, are known to show high critical current densities ($J_c$) of ~1 MA/cm$^2$, i.e., they have a strong vortex pinning ability (19–26). Based on Bean's critical state model (27), $J_c$ is proportional to the gradient of the magnetic flux density ($\mu_0 J_c = |dB/dx|$). Then, when an IBS with a sample width ($t$) of 0.1 mm is magnetized in an external field, a large number of vortices are pinned and a strong magnetic field ($H_{sc} = J_c t/2$) of ~5 kOe is estimated to be generated at the center of the sample. Therefore, in the case of Eu-IBSs, the superconductivity and the magnetic subsystem are both a source of strong magnetic fields; consequently, a non-trivial situation, where the magnetic fields generated by each state mutually affect each other, is considered. In this context, EuRbFe$_4$As$_4$ with the highest $T_c$ among Eu-IBSs is an optimal material to investigate the influence of superconductivity on magnetism.

Fig. 1$A$ shows the crystal structure of EuRbFe$_4$As$_4$. Since the Eu and Rb layers are alternately stacked without a solid solution, the Eu layer spacing is approximately twice that of EuFe$_2$As$_2$. Fig. 1$B$ illustrates the temperature dependence of the magnetic susceptibility, showing a sharp superconducting transition at $T_c$ = 37 K with perfect diamagnetism at zero-field cooling. The kink structure at $T_m$ = 15 K found below $T_c$ corresponds to the magnetic ordering of the Eu$^{2+}$

moments. The arrows in Fig. 1*A* indicate the direction of the $Eu^{2+}$ moments. In the ordered state, the $Eu^{2+}$ moments align ferromagnetically in the *ab* plane, and the orientation of the ferromagnetic alignment rotates by 90° in the next layer, i.e., the ground state is a helical order (12). Meanwhile, magnetization measurements showed that the $Eu^{2+}$ moments are re-oriented to the ferromagnetic alignment at relatively low external fields (the saturation field ($H_{sat}$) is estimated to be ~2 kOe) (28). In the ferromagnetically aligned state, the internal field ($H_{Eu}$) calculated from the $Eu^{2+}$ moments (7 $\mu_B$/Eu) is ~4.5 kOe, which is approximately half of that in $EuFe_2As_2$, reflecting the Eu layer spacing.

Thus, in $EuRbFe_4As_4$, the magnitudes of the characteristic magnetic fields related to superconductivity and magnetism, i.e., $H_{sc}$, $H_{Eu}$, and $H_{sat}$, are comparable to each other, such that the $Eu^{2+}$ moments cannot neglect the influence arising from superconductivity. In principle, it is possible to control the arrangement of local $Eu^{2+}$ moments using $H_{sc}$. Moreover, because the internal field generated by superconducting currents has spatial distributions, the direction of the $Eu^{2+}$ moments is expected to be locally manipulable by controlling the magnetic field profile. In this study, we demonstrate that (i) the orientation of the $Eu^{2+}$ moments in magnetic fields is dominated by the direction of pinned vortices, and (ii) the domain structure of the $Eu^{2+}$ moments can be manipulated by the distribution of superconducting vortices.

**Results and Discussion**

Fig. 2*A* shows the magnetization hysteresis loop (MHL) of $EuRbFe_4As_4$. The shape of the MHL can be interpreted as a superposition of the ferromagnetic contribution of the $Eu^{2+}$ moments on the hysteresis caused by vortex pinning of the superconductor. Assuming that the $Eu^{2+}$ moments do not exhibit hysteresis in the increasing and decreasing field processes (Fig. 2*B*), as the simplest two-component model, the hysteretic part of the MHL ($\Delta M = M_- - M_+$, Fig. 2*C*) depicts the contribution of the vortex pinning, and the average magnetization ($M_{ave} = (M_+ + M_-)/2$, Fig. 2*D*) depicts that of the $Eu^{2+}$ moments (28, 29). In fact, the hysteresis of the $Eu^{2+}$ moments in $EuFe_2As_2$ (a related non-superconducting material) is negligibly small (30), which likely justifies the assumption. However, this approach is too simplistic, as we show subsequently.

$M_{ave}$ saturates around $H_{sat}$ ~ 4 kOe, as shown in Fig. 2*D*. This indicates the re-orientation of the $Eu^{2+}$ moments from the helical structure to a ferromagnetic arrangement (regarding the difference in the magnitude of $H_{sat}$ from literature data (28), see Supplementary Information). The saturation magnetization of ~330 emu/cm$^3$ corresponds to 6.9 $\mu_B$/Eu, which is close to the expected full moment of 7 $\mu_B$/Eu. Meanwhile, based on Bean's critical state model (27), the magnetically-determined $J_c$ can be estimated from $\Delta M$, as shown in Fig. 2*E*. The magnitude of $J_c$ in the low-field region exceeds 1 MA/cm$^2$, which is comparable to those reported for 122- and 1144-type IBSs (19, 20, 24–26). This indicates that numerous vortices are pinned in the sample at zero external field after applying a large field. The magnitude of the trapped magnetic field at the center of the sample is estimated to be ~5 kOe, which is larger than $H_{sat}$ required for the ferromagnetic alignment of the $Eu^{2+}$ moments. Subsequently, as shown schematically in Fig. 2*F*, the pinned vortices significantly influence the $Eu^{2+}$ moments, i.e., neglecting any hysteresis of the $Eu^{2+}$ moments may be too simplistic.

In the present case, even if the $Eu^{2+}$ moments exhibit a ferromagnetic hysteresis, it is difficult to extract the hysteresis from the results of magnetization measurements due to the large magnetization hysteresis arising from the vortex pinning of superconductivity below $T_c$. To directly observe the behaviour of the $Eu^{2+}$ moments in the superconducting state, we investigated single-crystal neutron diffraction patterns of $EuRbFe_4As_4$ under magnetic fields.

Fig. 3*A* shows the neutron diffraction patterns along the (00*L*) direction of $EuRbFe_4As_4$ at 2.5 K, well below $T_m$, with the field applied along the *ab* plane. The open (solid) symbol data were measured under an increasing (decreasing) field. First, in the zero-field-cooling (ZFC) condition

(open blue), diffraction peaks are observed at (00$L\pm$0.25) positions in addition to (00$L$) positions. This (0, 0, 0.25) magnetic propagation vector indicates that the direction of the Eu$^{2+}$ moments rotates by 90° between neighbouring layers, i.e., the helical structure, as described above (Fig. 1$A$). With increasing magnetic field, at 1 kOe (open green) and 2 kOe (open orange), the intensity of the (00$L$ ± 0.25) peaks is suppressed (the enlarged view around the (001.25) peak is shown in Fig. 3$B$), while that of the (00$L$) peaks is enhanced. This change in the peak intensity demonstrates the transition of the Eu$^{2+}$ magnetic order from a helical structure to a ferromagnetic arrangement. With further increasing field, the (00$L$ ± 0.25) peaks almost disappear at 4 kOe (open red), and the diffraction pattern becomes identical to that observed at a large field of 2 T (open black). Thus, the Eu$^{2+}$ moments completely align ferromagnetically at ~4 kOe, which is comparable to $H_{sat}$ extracted from the MHL at 2 K.

Next, the field was decreased from 2 T. The diffraction pattern at 2 kOe (solid orange) is found to be almost identical to that at 20 kOe. At zero field, although the peak at (001.25) position can be identified, the intensity is approximately one tenth of that in the ZFC condition. The results clearly show that the Eu$^{2+}$ moments do not return to the helical structure, but remain in the ferromagnetic arrangement even at zero applied field. By further increasing the field in the negative direction, the (001.25) peak is slightly enhanced at −1 kOe (solid light green), and a similar diffraction pattern is observed at −2 kOe (solid brown). At −4 kOe (solid pink), the (001.25) peak is completely suppressed again, indicating the ferromagnetic arrangement of the Eu$^{2+}$ moments in negative direction.

The magnetic field dependence of the intensity of the (001.25) peak (blue) and the (001) peak (red), representing the fraction of helical and ferromagnetic structure, respectively, is summarized in Fig. 3$C$. The intensities of the (001) and (001.25) peaks are inversely correlated. Based on these results, the field dependence of the fraction of helical and ferromagnetic Eu$^{2+}$ moments is plotted in Fig. 3$D$. Here, for simplicity, no intermediate magnetic structure between the helical and the ferromagnetic structures is considered. This assumption is reasonable because the fraction of intermediate structure is estimated to be small (see Supplementary Information). Evidently, this behavior of the Eu$^{2+}$ moments in Fig. 3$D$ is distinct from the model shown in Fig. 2$B$, as only one part of the sample exhibits the helical structure in the field-decreasing process, and the region is not centered at zero, but appears in the negative field region. Fig. 3$E$ shows the magnetic field dependence of the magnetization arising from the Eu$^{2+}$ moments ($M_{Eu}$), calculated based on Fig. 3$D$. The calculated $M_{Eu}$ clearly shows a ferromagnetic hysteresis with the coercive field of ~1 kOe, which is two orders of magnitude larger than that of EuFe$_2$As$_2$. Therefore, the magnetization hysteresis ($\Delta M$) in the low-field region measured at low temperatures below $T_m$ must be considered as the sum of the hysteresis arising from the vortex pinning ($\Delta M_{SC}$) and that induced in the Eu$^{2+}$ moments ($\Delta M_{Eu}$).

To explore the ferromagnetic hysteresis of the Eu$^{2+}$ moments, we review the magnetic field dependence of $M$ in detail. Here, we focus on the MHLs in the low-field region (±10 kOe), where the hysteretic behaviour of the Eu$^{2+}$ moments is expected. Fig. 4$A$ shows representative MHLs measured below and above $T_m$. A clear hysteresis is observed at all temperatures, and the size of the MHL is largest for the lowest temperature (2 K, black) and decreases with increasing temperature. Below $T_m$, the magnitude of $M_-$ ($M_+$) under a positive (negative) field is almost constant, while the MHLs above $T_m$ are shrunk around zero field. In non-magnetic superconductors, the hysteresis is dominated by vortex pinning, and the width $\Delta M$ corresponds to the magnitude of $J_c$. Fig. 4$B$ shows the magnetic field dependence of $\Delta M$ at each temperature derived from Fig. 4$A$. At 2 K (black), $\Delta M$ shows a peak at zero field and decreases with increasing/decreasing field. This appears to be a typical magnetic field dependence of $J_c$. Further, a small kink structure (indicated by a black arrow) reflecting a jump in the MHL is observed around ±4 kOe. By increasing the temperature to 5 K (blue), 10 K (green), and 15 K (orange), $\Delta M$ decreases, and the kink position shifts to lower fields. By further increasing the temperature, at 20 K (red) above $T_m$, the peak observed at zero field changes to a dip, and the kink structure disappears. $\Delta M$ shows a similar field dependence with a smaller magnitude at higher temperatures (25 K (purple) and 30 K (light green)). This field dependence of $\Delta M$ ($J_c$), having a dip at zero field, is not common; however, similar behaviour was observed for the $\Delta M$ of CaKFe$_4$As$_4$ and associated with the defect structure specific to 1144-type materials (26). Fig. 4C shows the temperature and field dependence of $\Delta M$ plotted in a contour plot. $\Delta M$ peaks at zero field below $T_m$ = 15 K (horizontal dashed line) and dips at higher temperatures. The change in the field dependence of $\Delta M$ across $T_m$ was a strange behaviour, supposing it reflects that of $J_c$, while it is reasonable considering that the hysteresis of the Eu$^{2+}$ moments contributes to $\Delta M$ at low fields only below $T_m$. Fig. 4D shows the decomposition of $\Delta M$ at 2 K into the contribution of the Eu$^{2+}$ moments ($\Delta M_{Eu}$) and superconductivity ($\Delta M_{sc}$) using the magnetic field dependence of $M_{Eu}$ determined from neutron diffraction experiments (Fig. 3E). $\Delta M_{sc}$ (blue) has a dip at zero field similarly to those above $T_m$, indicating that the field dependence of $J_c$ does not change across $T_m$. Moreover, $\Delta M_{Eu}$ (red) becomes zero around ±4 kOe, which approximately coincides with the magnetic field, where a kink appears in $\Delta M$ (open triangles in Fig. 4C), suggestive of an interaction between those two.

Subsequently, we test whether these experimental results can be understood assuming that the state of the Eu$^{2+}$ moments (upward, helical, or downward) is determined by the local magnetic flux density ($B$) in the sample. Here, the change in the distribution of $B$ in the field-decreasing process from a sufficiently high field is considered. We separate the $B$ profile into those created by vortex pinning and the Eu$^{2+}$ moments, and model it.

Based on Bean's critical state model (27), the distribution of $B$ pinned in a superconductor exhibits a 'roof-top' shape, as shown in the left drawing in Fig. 5A. For simplicity, the field dependence of $J_c$ is not considered, i.e., the slope of the $B$ profile inside the sample (d$B$/d$x$) is taken as constant. The graphs in the right upper panel of Fig. 5A show the $B$ profiles of the roof-top shape cut along the red dashed line in the left drawing under different external fields. $B$ inside the sample is described as $B = \mu_0 H_{ex} + \mu_0 J_c x$, where $H_{ex}$ is the external field, and $x$ is the distance from the sample surface. At the center of the sample, $B$ becomes $\mu_0 H_{ex} + \mu_0 H_{sc}$, where $H_{sc} = J_c t/2$ and $t$ is the sample size along the cut. Setting $J_c$ = 0.8 MA/cm$^2$ (Fig. 2E) and $t$ = 0.13 mm, $H_{sc}$ is estimated to be 6.5 kOe. Since the field dependence of $J_c$ is ignored, the $B$ profile shifts downward without changing the shape with decreasing $H_{ex}$ (from (i) to (vi)).

Next, we model the $B$ profile arising from the Eu$^{2+}$ moments, supposing that the $B$ profile created by vortex pinning determines the arrangement of the Eu$^{2+}$ moments. For simplicity, we assume that the Eu$^{2+}$ moments order helically for $|B| < B_0$, where $B_0$ is a threshold field (the range shaded by pink in the upper panel of Fig. 5A) and align ferromagnetically in the upward or downward direction according to the sign of $B$ for $|B| > B_0$. Fig. 3D shows that the fraction of the helical structure reaches approximately 10 % in a decreasing field. Therefore, we assume $2B_0$ = 0.5 kG (approximately 10 % of $H_{sc}$), where the fraction of the helical structure is expected to be ~10 %. Thus, the $B$ profiles of the Eu$^{2+}$ moments at each $H_{ex}$ are derived, as shown in the lower panel of Fig. 5A. For $\mu_0 H_{ex} > B_0$ (i), the Eu$^{2+}$ moments maintain an upward ferromagnetic arrangement and the internal field ($H_{Eu}$) is 4.5 kOe at any position. With decreasing $H_{ex}$, when $B_0 > \mu_0 H_{ex} > -B_0$ (ii and iii), the Eu$^{2+}$ moments maintain an upward ferromagnetic arrangement around the center of the sample, whereas the helical structure ($H_{Eu}$ = 0) is adopted at the edge of the sample, where $|B| < B_0$. Subsequently, when $\mu_0 H_{ex} < -B_0$ (iv), the Eu$^{2+}$ moments take a downward (upward) ferromagnetic arrangement with $H_{Eu}$ = −4.5 kOe (4.5 kOe) at the sample edge (center). The helical structure appears at the region where $|B| < B_0$, which corresponds to the boundary between the upward and downward ferromagnetic domains. By further decreasing $H_{ex}$ (v), the downward (upward) ferromagnetic domain becomes larger (smaller), while the change in helical area is small, as it appears at the domain boundary. When $\mu_0 H_{ex} = B_0 − \mu_0 H_{sc}$ = −6.25 kG (vi), the upward ferromagnetic domain disappears. Consequently, the helical region sandwiched by the downward ferromagnetic domain only feels the negative $B$ (< −$B_0$), and thus, the whole region will suddenly turn into the downward ferromagnetic state (vii). This sudden disappearance of the helical region gives rise to the discontinuous change of *M*.

Based on the model, the magnetic field dependence of the $Eu^{2+}$ moment state is calculated, as shown in the left panel of Fig. 5*B*. The features are as follows: (1) the helical component appears around 0 Oe and is present down to –6 kOe (asymmetric with respect to zero field), and (2) the volume fraction of the helical component is approximately 10 % across the entire region. Comparing this with the neutron diffraction results (left panel of Fig. 5*C*), the magnetic field dependence of the upward ferromagnetic (black), helical (blue), and downward ferromagnetic (red) components is mostly reproduced. The right panel of Fig. 5*B* shows the calculated $M = M_{sc} + M_{Eu}$ where $M_{sc}$ is constant in this model, and $M_{Eu}$ is derived from the left panel of Fig. 5*B*. The *M* in the field-increasing process is calculated supposing $M_+(H) = -M_-(-H)$. The calculated MHL is in good agreement with the measured MHL shown in the right panel of Fig. 5*C*, including the magnitude and the kink structure. The slight deviation is considered to come from the simplification in the present model, i.e., the field dependence of $J_c$ and the intermediate state of $Eu^{2+}$ between the helical and ferromagnetic structure are not taken into account (see Supplementary Information).

These results support the model that the local *B* created by superconducting vortices determines the $Eu^{2+}$ spin domain structure in $EuRbFe_4As_4$. If this is the case, it is expected that various spin textures can be formed by changing the *B* profile. To provide an example, we attempt to introduce a larger number of ferromagnetic domains in the present $EuRbFe_4As_4$ sample. Here, we consider a process where $H_{ex}$ is switched from a decreasing to an increasing direction at some field ($H_{rev}$), as shown in Fig. 6*A*. Then, following Bean's critical state model, the *B* distribution changes from the sample edge, while it remains unchanged around the sample center. When $\mu_0 H_{ex} > B_0$, a domain structure shown in Fig. 6*C* is expected, where the sample center and edge are upward ferromagnetic (green), the region between them is downward ferromagnetic (blue), and the domain boundaries (grey) are helical. The left panel of Fig. 6*B* shows the calculated *M* – *H* curves based on the model with various $H_{rev}$ values of 0, –2, –5, and –10 kOe, and the right panel shows the measured *M* – *H* curves. The calculated results are in good agreement with the experiments, supporting the validity of the proposed model. For $H_{rev} = 0$ (blue), because the $Eu^{2+}$ moments maintain the upward ferromagnetic arrangement, the change in *M* reflects the change in the *B* profile associated with vortex pinning (*B* profile is reversed). Meanwhile, for $H_{rev} = -10$ kOe (red), because the $Eu^{2+}$ moments in the entire sample are aligned downward ferromagnetically, and the *B* profile by vortex pinning is almost reversed around $H_{ex} \sim 0$ in the field-increasing process; the situation is similar to Fig. 5*A* with an opposite sign. In contrast, in the case of $H_{rev} = -2$ and –5 kOe, which satisfy the condition of $\mu_0|H_{rev}| < \mu_0 H_{sc} - B_0$, the upward ferromagnetic domain remains at the sample center, and thus, the domain structure shown in Fig. 6*C* is expected to be realized for the field range of $B_0 < \mu_0 H_{ex} < \mu_0 |H_{rev}| - B_0$.

According to theoretical works (31, 32), the helical order of the $Eu^{2+}$ moments of $EuRbFe_4As_4$ (in zero field) appears due to the presence of superconductivity. It is proposed that the frustrating interlayer interactions caused by the interplay between the normal (ferromagnetic) and superconducting (antiferromagnetic) parts lead to the helical structure. When an external magnetic field is applied, each vortex generates a magnetic field decaying over a distance of the penetration depth ($\lambda$) from the core. For $EuRbFe_4As_4$, $\lambda$ is estimated to be 100-200 nm at low temperatures (16), which is much larger than the Eu-Eu layer distance of 1.3 nm. Accordingly, a direct magnetic interaction between vortices and the $Eu^{2+}$ moments would become dominant and thus the $Eu^{2+}$ moments favor the ferromagnetic alignment in the presence of vortices. Moreover, the present results show that the $Eu^{2+}$ moments must be ferromagnetically aligned by a much smaller field than that previously estimated from $H_{sat}$ because the experimental data were reproduced by setting $B_0$ to 0.25 kG, which is one order of magnitude smaller than $\mu_0 H_{sat}$. Notably, the average intervortex distance ($a_0$) at $B_0$ is calculated to be ~300 nm using the formula $B = 2\Phi_0/\sqrt{3}a_0^2$ for a triangular flux lattice where $\Phi_0$ is the magnetic flux quantum. The value of $a_0$ is close to $2\lambda$, indicating that the $Eu^{2+}$ moments align ferromagnetically when the whole region is covered by the field generated by vortices. Therefore, we consider $B_0$ to be the intrinsic saturation field of the $Eu^{2+}$ moments in $EuRbFe_4As_4$, while $H_{sat}$ extracted from $M_{ave} - H$ is a sample dependent parameter (see Supplementary Information).

Finally, the experiments in this work were performed with a magnetic field along the *ab* plane, which is the magnetic easy plane of $EuRbFe_4As_4$. When the magnetic field is applied along the *c* axis, the $Eu^{2+}$ moments are expected to tilt continuously from the *ab*-plane direction without destroying the helical structure. Meanwhile, similarly to the case of $H_{ex}$ // *ab*, the $Eu^{2+}$ moments are easily aligned along the *c* axis (28), and the superconducting properties of $EuRbFe_4As_4$ are rather three dimensional (33). Then, the present model is expected to be applicable for $H_{ex}$ // *c* by changing the parameters, $J_c$, $t$, and $B_0$, and modeling $M_{Eu}$ for the intermediate conical structure instead of using $B_0$ will improve the model. However, the situation is more complicated for $H_{ex}$ // *c* because the present sample is thin along the field direction which will result in significant demagnetization effects. In this case, the *B* profile inside the sample no longer follows that of the Bean's critical state model owing to a strong stray field, hence the model must be modified to describe the *B* profile including the stray field.

**Conclusions**

In summary, we revealed that the $Eu^{2+}$ moments exhibit a ferromagnetic hysteresis in the magnetic superconductor $EuRbFe_4As_4$, which is characterized by a high $T_c$ and $T_m$, as well as a strong vortex pinning ability. These experimental results can be explained by a model where the state of the $Eu^{2+}$ moments is dominated by the local *B* distribution created by vortex pinning. Using the unique interplay between the $Eu^{2+}$ moments and superconducting vortices, we demonstrated that the magnetic domain structure can be manipulated by controlling the *B* profile. We changed the *B* profile by switching the external field, and expect that the construction of more complex domain structures is possible by controlling the local *B*. Thus far, devices to manipulate local spins via superconducting vortices have been proposed for artificial superconductor/magnet hybrids (18). The present results demonstrate the possibility that such a device may be realized by utilizing $EuRbFe_4As_4$, which is a natural superconductor/magnet hybrid.

**Materials and Methods**

*Single-crystal growth*
Single crystals of $EuRbFe_4As_4$ were grown by the RbAs-flux method (34). EuAs, $Fe_2As$, and RbAs precursors were prepared from Eu and As, Fe and As, and Rb and As, respectively, which were mixed at appropriate molar ratios. The mixtures were sealed in evacuated quartz tubes (EuAs and $Fe_2As$) and a stainless steel tube with an alumina crucible (RbAs) and heated at 750 ℃ (EuAs), 900 ℃ ($Fe_2As$), and 600 ℃ (RbAs) for 20 h. EuAs, $Fe_2As$, and RbAs were weighed at a ratio of 1: 1: 15 to yield a total amount of 9 g and placed in a alumina crucible, then sealed in a stainless-steel container (35, 36). The container was heated for 5 h to 700 ℃ and maintained at this temperature for 5 h. It was then heated to 970 ℃ within 5 h and maintained this temperature for 10 h. Subsequently, it was cooled to 620 ℃ for 350 h (1 ℃/h). For single crystals used in this study, X-ray diffraction (XRD) patterns were measured at room temperature using a diffractometer with Cu Kα radiation (Rigaku, Ultima IV) to verify the presence of 00*l* peaks (see Supplementary Information). The as-grown crystals show a trace of $RbFe_2As_2$ (34); hence, the surfaces of as-grown crystals were carefully removed by cleaving. The cleaved surfaces only show 00*l* peaks from $EuRbFe_4As_4$.

*Magnetization measurements*

The samples for magnetization measurements were cut into rectangular shapes. The dimensions of the EuRbFe$_4$As$_4$ crystal used in this study were $l$ = 1.2 mm, $w$ = 0.7 mm, and $t$ = 0.13 mm, with the shortest edge along the $c$ axis. Measurements were performed using a magnetic property measurement system (Quantum Design). The magnetic field was applied along the $ab$ plane (along the $w$ edge), therefore two $J_c$ components (in-plane $J_c$ ($J_c^{ab}$) and inter-plane $J_c$ ($J_c^c$)) contribute to $M$. For Fig. 2*E*, we used a simplified formula for the evaluation of $J_c$ by taking $J_c^{ab} = J_c^c$, i.e., $J_c$ = 20Δ$M$/$t$(1 − $t$/3$l$) where Δ$M$ is the hysteretic part of the MHL and $l$ and $t$ are the sample dimensions in units of emu/cm$^3$ and cm, respectively. For the calculation of $M$ in Figs. 5 and 6, the anisotropy of $J_c$ ($J_c^{ab}/J_c^c$) was taken into account based on the evaluation of $J_c^{ab}$ and $J_c^c$ using an extension of Bean's critical state model for anisotropic $J_c$ (24) (see Supplementary Information).

*Neutron diffraction measurements*

Single-crystal neutron diffraction measurements on EuRbFe$_4$As$_4$ were carried out using the time-of-flight (TOF) single-crystal neutron diffractometer SENJU (37) installed at the Materials and Life Science Experimental Facility (MLF), Japan Proton Accelerator Research Complex (J-PARC). The sample size was 1.8 × 0.9 × 0.08 mm$^3$. A neutron wavelength of 0.4–4.4 Å was used. The magnetic field was applied along the [110] direction up to 20 kOe. The data was visualized by the software STARGazer (38).

**Acknowledgments**

This work was supported by the Austria-Japan Bilateral Joint Research Project hosted by the Japan Society for the Promotion of Science (JSPS) and by FWF: I2814-N36, and a Grant-in-Aid for Scientific Research (KAKENHI) (JSPS Grant No. 19K15034, 19H05823 and JP16H06439). The neutron diffraction experiment at SENJU was conducted under the user program of Proposal No. 2019B0066. We thank Yoshiyuki Yoshida for technical assistance.

**Figures**

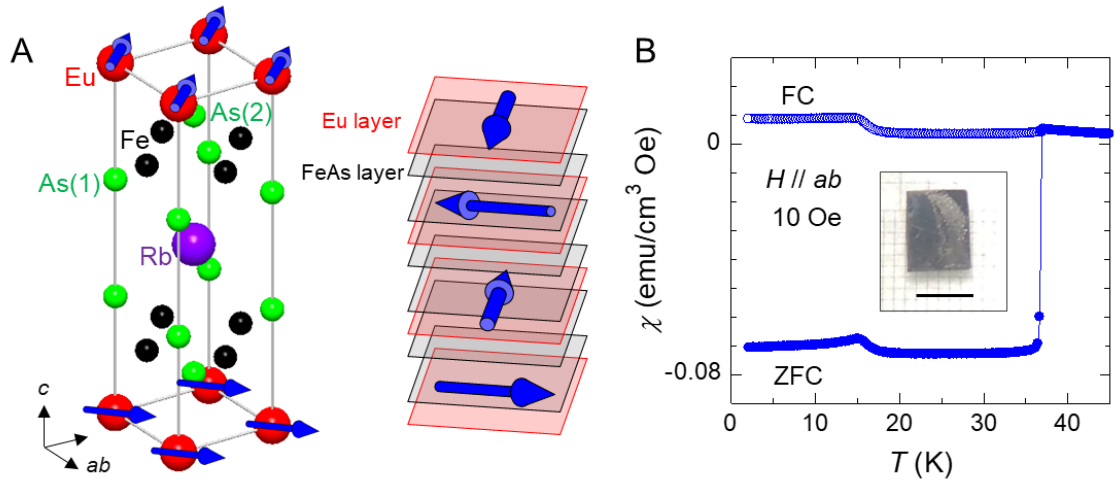

**Figure 1.** Crystal structure and magnetic ordering of EuRbFe$_4$As$_4$. (*A*) Crystal structure of EuRbFe$_4$As$_4$. Blue arrows indicate the direction of the Eu$^{2+}$ moments. The schematic diagram of the helical magnetic structure is shown on the right. (*B*) Temperature dependence of the magnetization of a EuRbFe$_4$As$_4$ single crystal measured for zero-field-cooling and field-cooling processes. A magnetic field of 10 Oe was applied along the *ab* plane. The inset shows a photograph of a EuRbFe$_4$As$_4$ single crystal with a black scale bar corresponding to 1 mm.

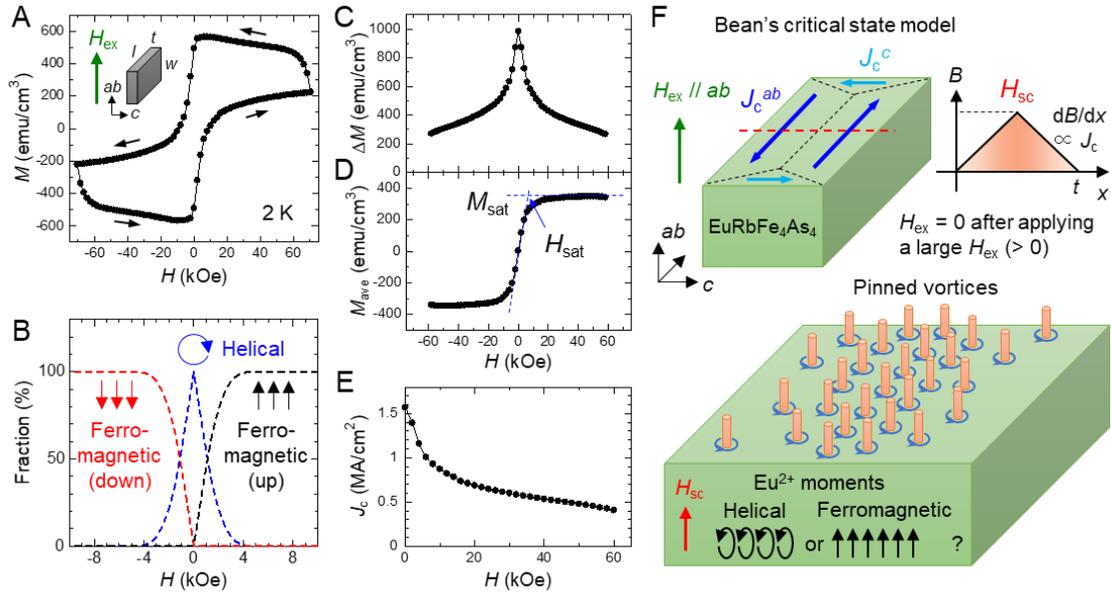

**Figure 2.** Magnetization of EuRbFe$_4$As$_4$ and supposed response of the Eu$^{2+}$ moments to external field. (*A*) Magnetization hysteresis loop of EuRbFe$_4$As$_4$ measured at 2 K with a magnetic field parallel to the *ab* plane. The inset shows the geometric alignment. Black arrows indicate the field sweep direction. (*B*) Response of the Eu$^{2+}$ moments to external magnetic field assuming the absence of hysteresis. The blue dashed line indicates the fraction of the helical structure, and the black (red) dashed line indicates that of the ferromagnetic structure in a positive (negative) applied field. (*C*) Hysteretic part of MHL, $\Delta M = M_- - M_+$, where $M_+$ ($M_-$) is the magnetization measured in the field-increasing (decreasing) process, derived from *A*. (*D*) Average magnetization $M_{ave} = (M_+ + M_-)/2$. The saturation field $H_{sat}$ was determined by the crossing point of the extrapolation lines from the slope at the low-field region and the saturation moment $M_{sat}$. (*E*) Magnetic field dependence of $J_c$ calculated using the formula $J_c = 20\Delta M/t(1 - t/3l)$, where *t* and *l* are the sample dimensions (*t* < *l*). Here, for simplicity, the anisotropy of $J_c$ flowing along the *ab* plane ($J_c^{ab}$) and *c* axis ($J_c^c$) is not considered, i.e. $J_c = J_c^{ab} = J_c^c$. (*F*) Schematic diagrams of the superconducting currents and the distribution of vortices in EuRbFe$_4$As$_4$. The top left drawing shows the superconducting currents flowing in the critical state at zero field after applying a large field. The top right graph shows the profile of the magnetic flux density along the red dashed line in the top left drawing. The magnitude of the magnetic field generated at the center of the sample is estimated to be $H_{sc} = J_c t/2$. The bottom drawing shows a schematic diagram of the distribution of vortices. Each orange rod surrounded by a blue clockwise arrow indicates a vortex composed of circular supercurrents generating an upward magnetic field.

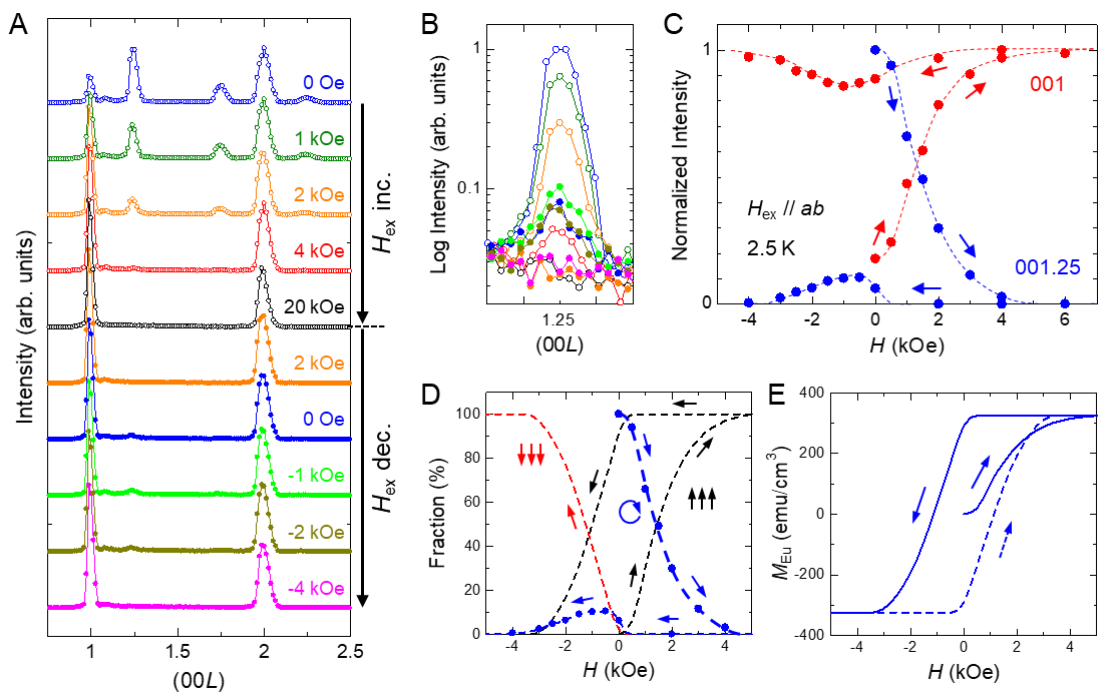

**Figure 3.** Field dependence of neutron diffraction patterns of EuRbFe$_4$As$_4$ and response of the Eu$^{2+}$ moments to external field. (*A*) Neutron diffraction patterns of EuRbFe$_4$As$_4$ along (00*L*) at 2.5 K under magnetic fields along *ab* plane. The pattern on the top (open blue circles) was measured after zero-field cooling, and the magnetic field was increased to 20 kOe from the top to middle (open black circles). From the middle to bottom, the field was decreased to –4 kOe (pink solid circles). (*B*) Magnified view around (001.25) peak plotted in log scale. (*C*) Magnetic field dependence of peak intensity for (001) peak (red) and (001.25) peak (blue). The peak intensity is normalized at 20 kOe and 0 Oe (ZFC) for the (001) and (001.25) peaks, where the Eu$^{2+}$ moments are completely helical and ferromagnetic, respectively. Arrows indicate the field sweep direction. (*D*) Fraction of the Eu$^{2+}$ moments assuming helical (blue) and ferromagnetic structures along positive (black) and negative (red) field directions, estimated from the results shown in *C*. Here, it is assumed that the upward ferromagnetic moment is aligned to the downward one via the helical structure for the *H* < 0 region in the field-decreasing process. (*E*) Field dependence of magnetization arising from the Eu$^{2+}$ moments ($M_{Eu}$), calculated based on *D*. The dashed line corresponding to $M_{Eu}$ in the field-increasing process from the negative field is derived from the field-decreasing data assuming $M_{Eu}(H) = -M_{Eu}(-H)$.

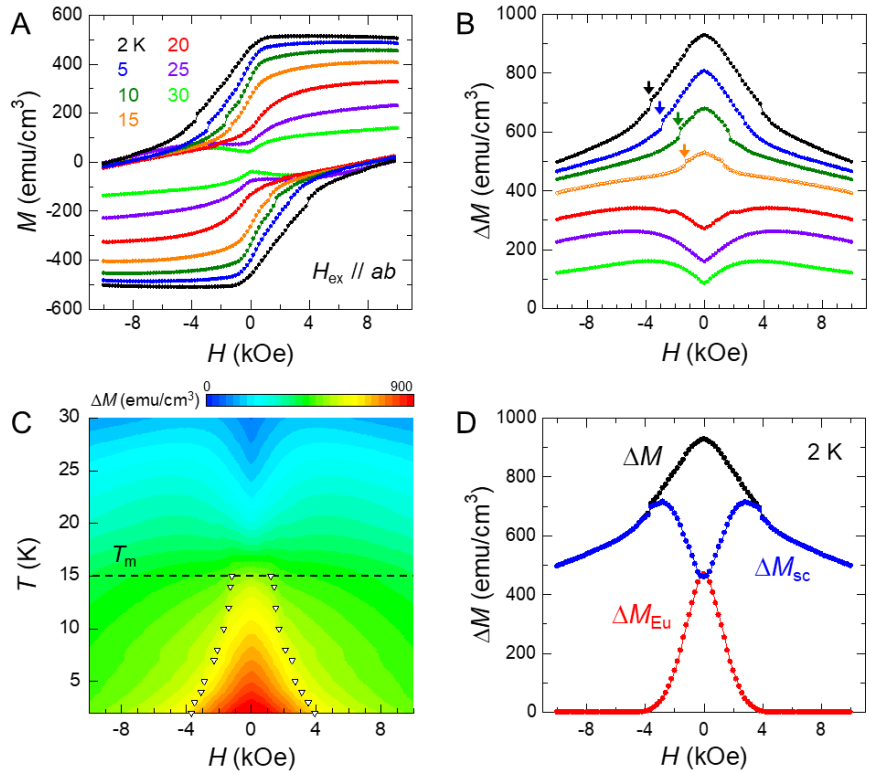

**Figure 4.** Magnetization hysteresis loops of EuRbFe$_4$As$_4$ at low field region. (*A*) MHLs under low fields measured at several temperatures below and above $T_m$ (2-30 K). Note that the field was applied along the *ab* plane up (down) to 30 (–30) kOe, whereas only data in the field range of ±10 kOe is shown. (*B*) Field dependence of $\Delta M$ derived from *A*. The same color code is used for each temperature as in *A*. Arrows indicate the kink structure, where *M* shows a sudden change. (*C*) Temperature and field dependence of $\Delta M$ plotted by a contour plot. The plot is produced using MHLs measured every 1-2 K. The hot (cold) colored region indicates that $\Delta M$ is large (small). The horizontal dashed line indicates $T_m$ (= 15 K). The open triangles correspond to positions of the kink structure. (*D*) Decomposition of $\Delta M$ at 2 K (black) into contributions from vortex pinning ($\Delta M_{sc}$, blue) and from the Eu$^{2+}$ moments ($\Delta M_{Eu}$, red). $\Delta M_{Eu}$ is derived from Fig. 3*C* and $\Delta M_{sc} = \Delta M - \Delta M_{Eu}$ is assumed to extract $\Delta M_{sc}$.

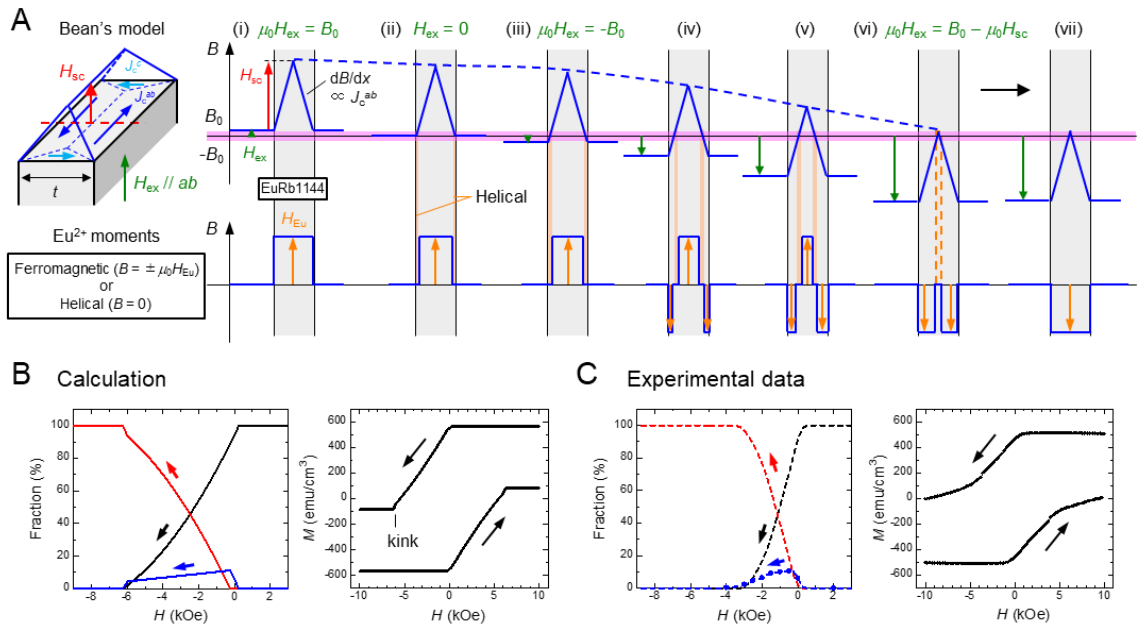

**Figure 5.** Model of flux density profile in EuRbFe$_4$As$_4$ and comparison of calculation and experimental results. (*A*) Change in *B* profile in field-decreasing process from a sufficiently high field. The left drawing shows a typical roof-top *B* distribution (indicated by blue lines) in superconductors expected from Bean's critical state model. Blue (light blue) arrows indicate the superconducting currents flowing along the *ab* plane (*c* axis). The magnetic field ($H_{sc}$) generated by the currents is indicated by the red arrow. The graphs in the right upper panel show *B* profiles cut along the red dashed line in the left drawing under different external fields. The horizontal pink shaded region indicates the range of $|B| < B_0 = 0.25$ kG, where the Eu$^{2+}$ moments assume the helical structure. Vertical orange shaded regions indicate the corresponding helical domains. The graphs in the right lower panel show the *B* profiles created by the Eu$^{2+}$ moments. The magnitude of the internal field is assumed to be $H_{Eu} = 4.5$, 0, and $-4.5$ kOe for upward ferromagnetic, helical, and downward ferromagnetic structures, respectively. (*B*) Calculation results of magnetic field dependence of fraction of upward ferromagnetic (black), helical (blue), and downward ferromagnetic (red) Eu$^{2+}$ moments (left panel) and total magnetization $M = M_{Eu} + M_{sc}$ (right panel). The arrows indicate the field sweep direction. (*C*) Experimental data extracted from Fig. 3d (left panel) and Fig. 4*A* (right panel). The same color code is used as in *B*.

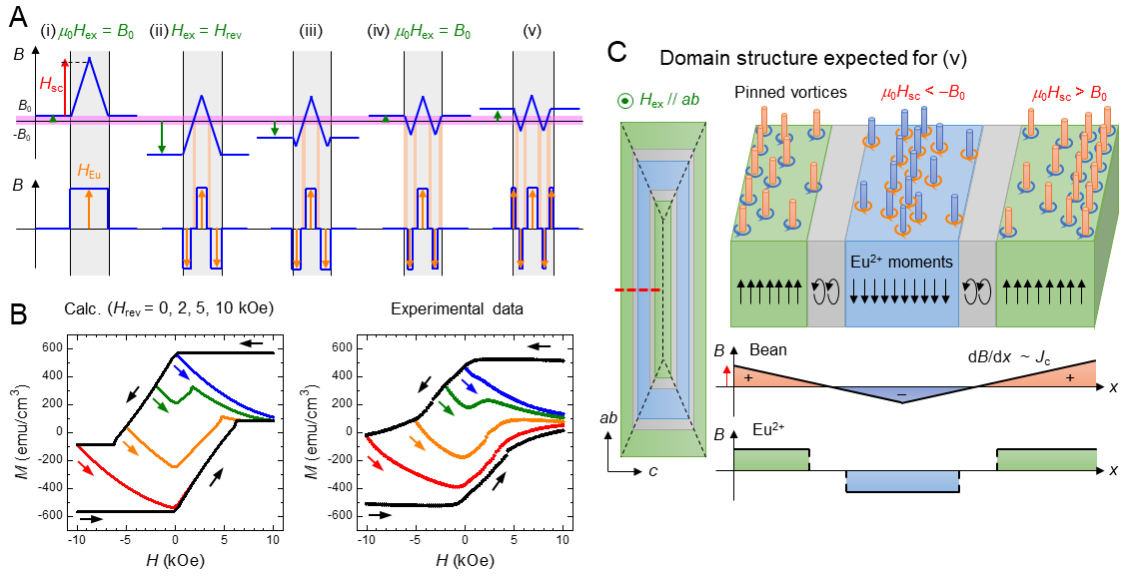

**Figure 6.** Demonstration of ferromagnetic domain control via trapped vortices in EuRbFe$_4$As$_4$. (*A*) Change in *B* profile with varying external field. The external field was first decreased (from (i) to (ii)) and then increased (from (ii) to (v)). (*B*) Calculated (left panel) and measured (right panel) magnetization curves for various $H_{rev}$ = 0 (blue), 2 (green), 5 (orange), and 10 kOe (red). The arrows indicate the field sweeping direction. (*C*) Eu$^{2+}$ domain structure expected for graph (v) in *A*. Vortices along positive (negative) field direction are colored orange (blue). Green, grey, and blue regions correspond to upward ferromagnetic, helical, and downward ferromagnetic domains, respectively.

## Supplementary Information for

Superconductivity-driven ferromagnetism and spin manipulation using vortices in the magnetic superconductor EuRbFe$_4$As$_4$

Shigeyuki Ishida[1*], Daniel Kagerbauer[2], Sigrid Holleis[2], Kazuki Iida[3], Koji Munakata[3], Akiko Nakao[3], Akira Iyo[1], Hiraku Ogino[1], Kenji Kawashima[4], Michael Eisterer[2] and Hiroshi Eisaki[1].

[1] Research Institute for Advanced Electronics and Photonics, National Institute of Advanced Industrial Science and Technology, Tsukuba, Ibaraki 305-8568, Japan

[2] Atominstitut, TU Wien, Stadionallee 2, 1020 Vienna, Austria

[3] Neutron Science and Technology Center, Comprehensive Research Organization for Science and Society (CROSS), Tokai, Ibaraki 319-1106, Japan

[4] Research & Development Department, IMRA JAPAN CO., LTD., Kariya, Aichi 448-8650, Japan

**Characterization of EuRbFe$_4$As$_4$ single crystals.** Prior to the magnetization and neutron diffraction measurements, the quality of the EuRbFe$_4$As$_4$ (EuRb1144) single crystals was verified by X-ray diffraction (XRD) measurements. Supplementary Fig. S1 shows XRD patterns of a EuRb1144 sample after cleaving the surfaces, which are presented on a linear (S1*A*) and a log scale (S1*B*). All peaks are indexed by 00*l* for EuRb1144. No traces of EuFe$_2$As$_2$ and RbFe$_2$As$_2$ phases were observed within the resolution of in-lab XRD apparatus.

**Calculation of $J_c$ based on an extended critical state model.** In Fig. 2 of the manuscript, we evaluated $J_c$ using a simplified formula which assumes $J_c^{ab} = J_c^c$. However, we also evaluated $J_c$, considering two $J_c$ components ($J_c^{ab}$ and $J_c^c$) based on an extension of Bean's critical state model for anisotropic $J_c$ (S1). We performed two independent $M - H$ measurements on a rectangular sample with dimensions *l*, *w* (< *l*), and *t* (// *c*) using different orientations, as illustrated in Supplementary Fig. S2*A*. Here, we show the calculation results for a EuRb1144 sample with dimensions *l* = 1.28 mm, *w* = 0.650 mm, and *t* = 0.156 mm. Supplementary Fig. S2*A* shows the two MHLs, $M_1$ and $M_2$, obtained from the two different orientations. Clearly, $\Delta M_1$ and $\Delta M_2$ differ from each other ($\Delta M_1 > \Delta M_2$). Supplementary Fig. S2*B* shows the obtained $J_c^{ab}$ (blue) and $J_c^c$ (red) using the calculation procedure described in previous studies (S2, S3). Moreover, the anisotropy of $J_c$, $J_c^{ab}/J_c^c$, is shown (black) in this figure. Note that these $J_c$ values are overestimated in the low-field region ($H < \sim6$ kOe) due to the hysteresis induced in the Eu$^{2+}$ moments ($\Delta M_{Eu}$), as discussed in the manuscript (see Fig. 4*D*).

**Estimation of the fraction of helical and ferromagnetic Eu$^{2+}$ moments.** In this work, no intermediate magnetic structure between the helical and the ferromagnetic structures was considered, i.e. we assumed that the total of helical and ferromagnetic Eu$^{2+}$ moments is 100 %. However, one would expect that some inhomogeneous magnetic structures appear under a magnetic field, which would not be observed by the neutron diffraction measurements but give a field-dependent contribution to the magnetization. Therefore, the existence of such intermediate structures would affect the calculation results and the conclusions if the contribution is significant.

Accordingly, we estimated the fraction of ferromagnetic $Eu^{2+}$ moments based on the change of the (001) intensity. Here, the intensity at 0 Oe (ZFC condition) corresponding to the nuclear Bragg peak was subtracted from the other data, and then normalized at 20 kOe where the $Eu^{2+}$ moments are considered to be ferromagnetically aligned completely. In Supplementary Fig. S3*A*, the field dependence of the fraction of helical and ferromagnetic $Eu^{2+}$ moments is plotted, showing the inverse correlation of the two fractions. In Supplementary Fig. S3*B*, the total value of the two fractions is plotted. It is demonstrated that the total fraction is approximately 100 % for the field-increasing process and 95 % for the field-decreasing process. Thus, the fraction of intermediate structures is estimated to be up to ~5 %. This would give a slight contribution to the magnetization, while it should not significantly affect the calculation results and thus does not alter the conclusions.

**Calculation of the magnetization with anisotropic $J_c$.** In Figs. 5 and 6 of the manuscript, for simplicity, the *B* profiles cut along one direction (short edge) are shown to describe the model. However, the *B* profiles along the other direction (long edge) must be taken into account to appropriately calculate the fraction of the helical domain for the sample with finite length (*l*). Moreover, the present $EuRbFe_4As_4$ sample shows an anisotropic $J_c$; therefore, the slope of the *B* profile (d*B*/d*x*) is different for the two directions, as shown in the top panels of Supplementary Fig. S4 ($J_c$ anisotropy ($J_c^{ab}/J_c^c$) is taken as 1, 5, and 10 for *A*, *B*, and *C*, respectively). Consequently, as shown in the middle panels of Supplementary Fig. S4, the width of the region in the range of $|B| < B_0$, which determines the helical fraction, is different for each direction depending on the $J_c$ anisotropy. The calculation results (bottom panels of Supplementary Fig. S4) show that the helical fraction is almost field-independent when $l/t \gg J_c^{ab}/J_c^c$, whereas it decreases with a decreasing field below $H = 0$ when $J_c^{ab}/J_c^c$ becomes larger. This decrease in the helical fraction is in good agreement with the neutron diffraction experiment results (Fig. 3*A*). As shown in the previous section, $J_c^{ab}/J_c^c$ of $EuRbFe_4As_4$ is approximately 5 for *H* values below 10 kOe; therefore, we chose $J_c^{ab}/J_c^c = 5$ (corresponding to Supplementary Fig. S4*B*) for the calculation of *M* in Figs. 5 and 6.

**Sample size dependence of the magnetization hysteresis and average magnetization.** The present study shows that the peak in $\Delta M - H$ below $T_m$ can be associated with a hysteresis of the $Eu^{2+}$ moments ($\Delta M_{Eu}$), and the calculated $M - H$ curves based on the proposed model are in good agreement with the results of the magnetization experiments. To further test the model, magnetization measurements were performed on samples with different sizes. Table S1 shows the dimensions of the samples #1-#3 and the calculated $H_{sc}$, assuming the same $J_c$. Notably, $H_{sc}$ is proportional to *t* in the present model, where d*B*/d*x* (~ $J_c^{ab}$) is constant, and $M_{Eu}$ must exhibit a hysteresis in the field range of $-(\mu_0 H_{sc} - B_0) < \mu_0 H_{ex} < \mu_0 H_{sc} - B_0$ (see Fig. 5). Supplementary Fig. S5*A* shows $\Delta M - H$ for the samples #1-#3 measured at 5 K with *H* along the *ab* plane. The magnitude of $\Delta M$ is similar for each sample around $H_{ex} = \pm 10$ kOe, while $\Delta M$ of sample #2 with the largest *d* (blue) becomes larger than the others below $|H_{ex}| <$ ~8 kOe. Subsequently, $\Delta M$ of sample #1 (red) starts to increase below $|H_{ex}| <$ ~6 kOe, and that of sample #3 with the smallest *t* (black) increases below $|H_{ex}| <$ ~4 kOe. The positions where $\Delta M$ indicates an increase roughly coincide with $H_{sc}$ indicated by the dashed lines in Supplementary Fig. S5*A*. These results support our model where the hysteresis of the $Eu^{2+}$ moments is induced by the self-field generated by the superconducting vortices. Similarly, the saturation field ($H_{sat}$) in $M_{ave} - H$ also depends on the sample size. Supplementary Fig. S5*B* shows $M_{ave} - H$ curves for samples #1-#3. One can see that the slope of $M_{ave} - H$ is larger (smaller) for sample #3 (#1) with a smaller (larger) *t*, resulting in a smaller (larger) $H_{sat}$ ~ 2 kOe (4 kOe). Note that $H_{sat}$ ~ 2 kOe for sample #3 with *t* = 0.083 mm is close to the literature data ($H_{sat}$ ~ 2 kOe) where *t* of the sample is 0.060 mm (S4). These results further support our model and explain the different $H_{sat}$ values observed in previous works.

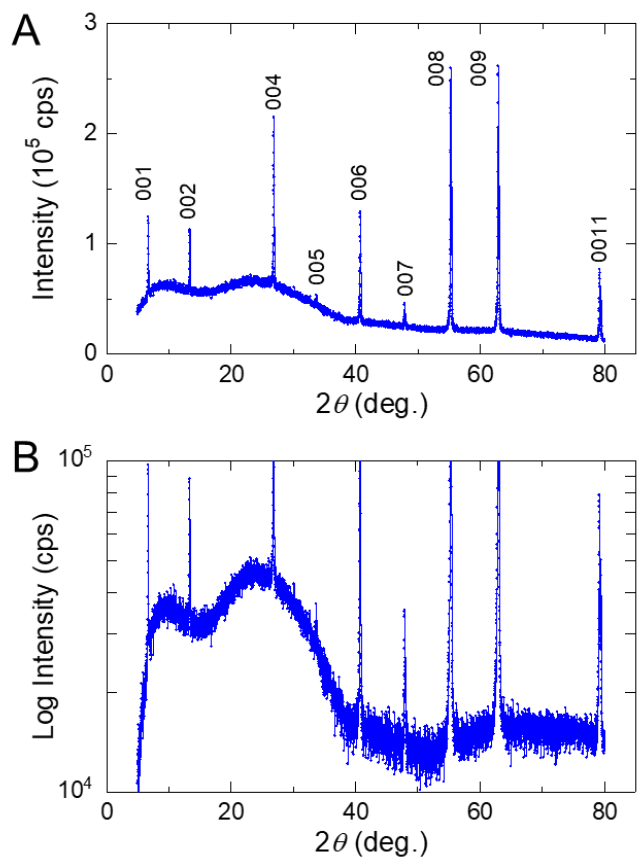

**Fig. S1.** XRD patterns for a EuRb1144 single crystal. The XRD patterns are presented on a linear (*A*) and a log scale (*B*), showing no traces of $EuFe_2As_2$ and $RbFe_2As_2$ phases.

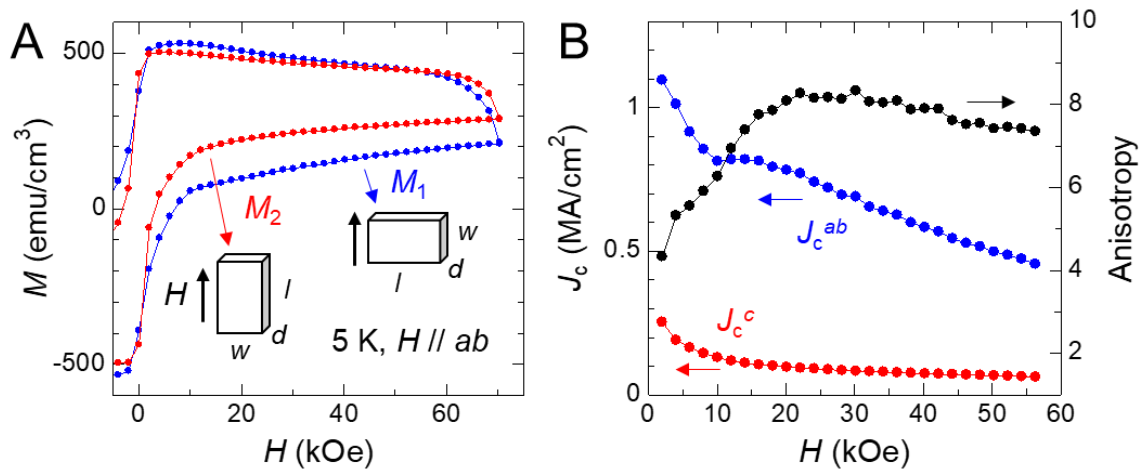

**Fig. S2.** Evaluation of $J_c^{ab}$ and $J_c^c$ for EuRb1144. (*A*) MHLs for two orientations ($M_1$ (blue, $H \mathbin{/\mkern-5mu/} w$) and $M_2$ (red, $H \mathbin{/\mkern-5mu/} l$)) at 5 K under $H \mathbin{/\mkern-5mu/} ab$. (*B*) $H$ dependence of $J_c^{ab}$ (blue) and $J_c^c$ (red) derived from $M_1$ and $M_2$ using an extension of Bean's critical state model. The anisotropy of $J_c$ defined as $J_c^{ab}/J_c^c$ is plotted in black.

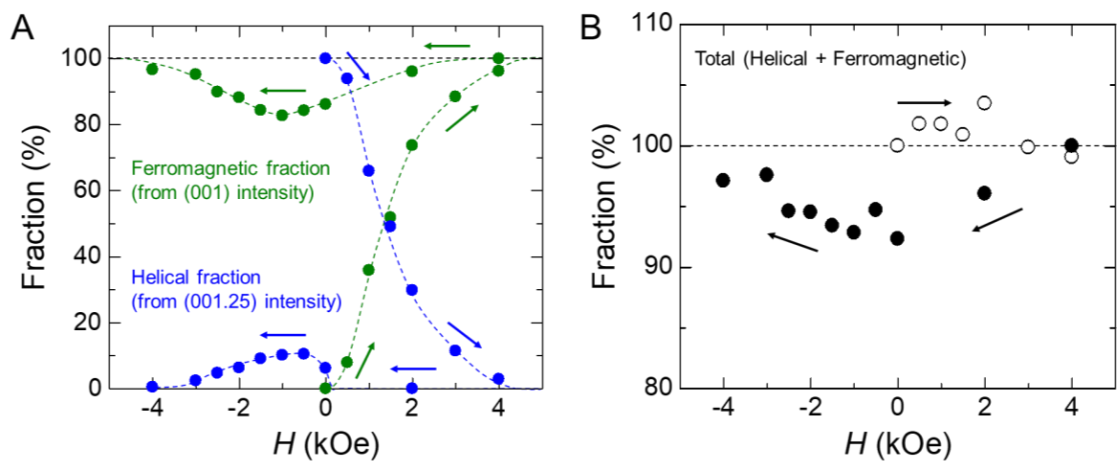

**Fig. S3.** Estimation of the fraction of helical and ferromagnetic $Eu^{2+}$ moments. (*A*) Field dependence of the fraction of helical (blue) and ferromagnetic (green) $Eu^{2+}$ moments estimated from the intensity of the (001.25) peak and the (001) peak, respectively. The arrows indicate the field sweep direction. (*B*) Total fraction of helical and ferromagnetic $Eu^{2+}$ moments. Open (solid) circles were measured under field-increasing (decreasing) process. The arrows indicate the field sweep direction.

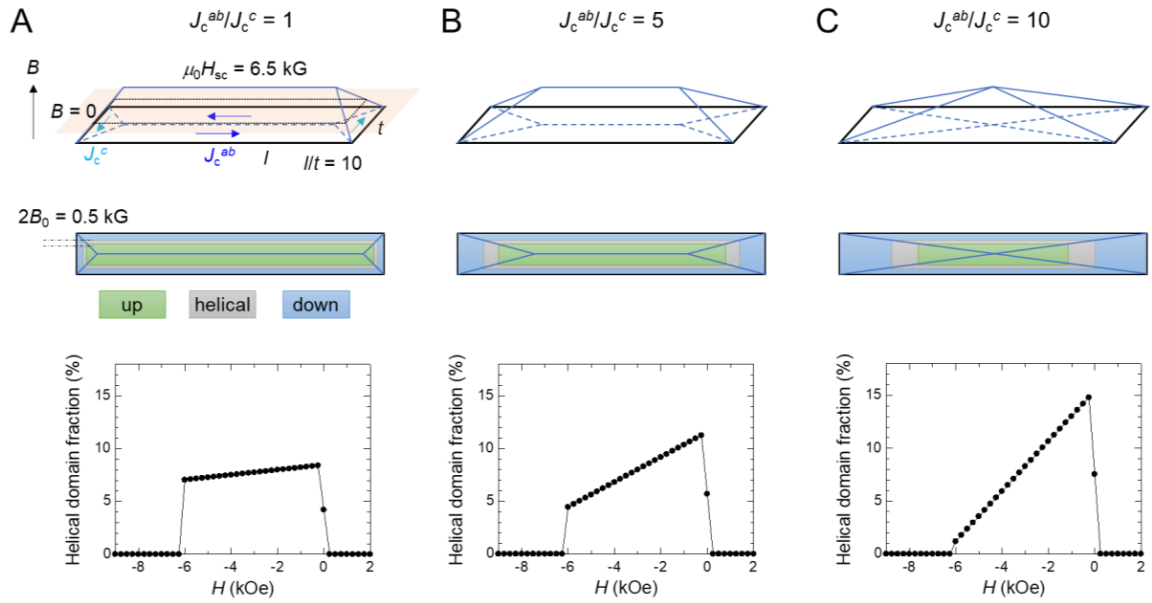

**Fig. S4.** Calculation of the helical domain fraction in decreasing field for different values of the $J_c$ anisotropy. (*A*) Isotropic $J_c$ ($J_c^{ab}/J_c^c = 1$). The top panel shows the *B* distribution. The ratio between the long edge (*l*) and short edge (*t*) is taken as $l/t = 10$, which is a typical value for EuRbFe$_4$As$_4$ samples used in this study. The light-orange plane indicates the position where $B = 0$. The middle panel illustrates the Eu$^{2+}$ spin domain structure. Green, grey, and blue regions correspond to upward ferromagnetic, helical, and downward ferromagnetic domains, respectively. The bottom panel depicts the field dependence of the helical domain fraction. (*B*) $J_c^{ab}/J_c^c = 5$. (*C*) $J_c^{ab}/J_c^c = 10$ (= $l/t$).

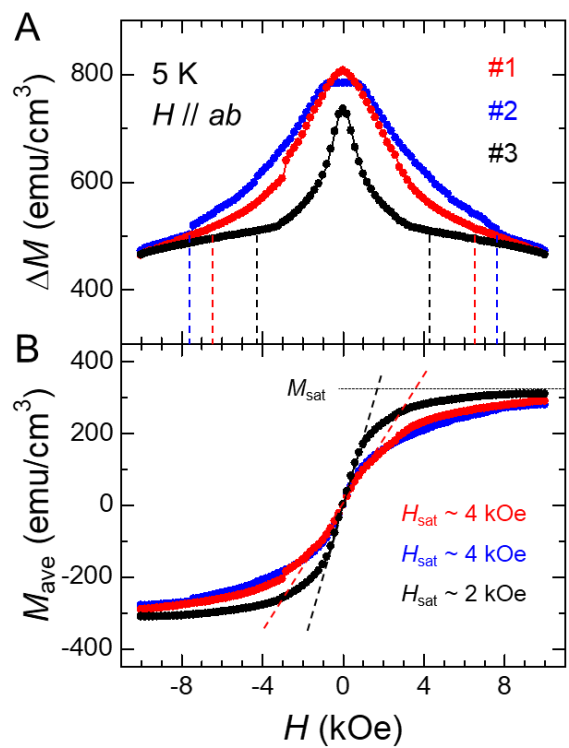

**Fig. S5.** Magnetization hysteresis for samples with different sizes. (*A*) $\Delta M - H$ measured at 5 K under *H // ab* for samples #1 (red), #2 (blue), and #3 (black). Dashed lines indicate the magnitude of the self-field ($H_{sc}$) estimated for each sample. (*B*) Corresponding $M_{ave} - H$. Dashed lines indicate the slope of $M_{ave} - H$ used to estimate $H_{sat}$.

**Table S1.** Dimensions of three samples and estimated self-field ($H_{sc}$) at the center of each sample.

| Sample | $l$ (mm) | $t$ (mm) | $l/t$ | $H_{sc}$ (kOe) |
|--------|----------|----------|-------|----------------|
| #1     | 1.20     | 0.130    | 9.23  | 6.5            |
| #2     | 1.28     | 0.156    | 8.31  | 7.8            |
| #3     | 1.29     | 0.083    | 15.5  | 4.2            |

**SI References**